\documentclass[submission,copyright,creativecommons]{eptcs}
\providecommand{\event}{ICLP 2019} 
\usepackage{underscore}           
\usepackage{listings}
\usepackage{dsfont}

\title{Solving a Flowshop Scheduling Problem with Answer Set Programming:
Exploiting the Problem to Reduce the Number of Combinations}
\author{Carmen Leticia Garc\'{i}a-Mata and Pedro Rafael M\'{a}rquez-Guti\'{e}rrez  
\institute{Tecnol\'{o}gico Nacional de M\'{e}xico - Tecnol\'{o}gico de Chihuahua\\
Chihuahua, Chih., M\'{e}xico}
\email{clgarcia; pmarquez@itch.edu.mx}
}
\def\titlerunning{Solving a Flowshop Scheduling Problem with Answer Set
Programming}
\def\authorrunning{C.L. Garc\'{i}a-Mata \& P.R. M\'{a}rquez-Guti\'{e}rrez}
\begin{document}
\maketitle

\begin{abstract}
Planning and scheduling have been a central theme of research in computer
science. In particular, the simplicity of the theoretical approach of a
no-wait flowshop scheduling problem does not allow to perceive the problem's
complexity at first sight. In this paper the applicability of the Answer Set
Programming language is explored for the solution of the Automated
Wet-etching scheduling problem in Semiconductor Manufacturing Systems. A
method based in ranges is proposed in order to reduce the huge number of
combinations.
\end{abstract}

\section{Introduction}

A distinctive characteristic of combinatorial problems is their massive
search space. This huge domain is due to the number of possible solutions
that although finite, grows exponentially with the amount of data. Some
typical combinatorial problems are the search for the cheapest or shortest
paths, internet data packets routing, protein structure prediction, and
planning and scheduling of resources.

In theory it is possible to find the optimal solution for each combinatorial
problem by conducting an exhaustive search. However, in practice finding an
optimal solution is often an intractable problem, even for problems of
modest size.

In this paper,  Answer Set Programming (ASP) is used to explore how to solve
the scheduling problem for an Automated Wet-etch Station (AWS) of a
Semiconductor Manufacturing System where the optimization objective is the
makespan. If a robot is not used to transfer jobs  between baths, the
problem can be approximated as a special case of the most general no-wait
scheduling flowshop problem. A flowshop is a multi-stage production process
where all jobs must pass through the same stages. There is a set  $J$ of
jobs with  $|J|\in \mathds{N}$ jobs in total.  All jobs must be processed
sequentially, starting on machine one and following a chain of $m$
operations $O_{j},\cdots,O_{jm}$. The operation $O_{ji}$ must be processed
for $t_{ji}$ time units at each stage. Each job is processed  at most by one
machine and each machine executes at most one job. Each operation is
performed without interruption in the machine that was assigned. The no-wait
constraint in scheduling problems occurs when two consecutive operations of
a job must be processed without interruptions. AWS's wafer etching consists
in processing $n$ jobs through a series of chemical and water baths
alternately ordered. Processing in chemical baths should immediately follow
the previous operation because an over exposition in the chemical bath can
damage the circuit. Once the wafer is processed in a water bath, it can be
used as local storage if needed.

According to  Graham's notation, this problem belongs to the
$F_m/no-wait/C_{max}$ class. A good survey about no-wait flowshop scheduling
problems is the one by Allahverdi \cite{article3}.

The three most popular AI methods used on the flowshop and AWS scheduling
problem are: a) meta-heuristics, like genetic algorithms (GA); particle
swarm optimization (PSO); ant colony optimization (ACO); Tabu Search (TS);
Simulated Annealing (SA); b) statistics: Monte Carlo methods and neural
networks; and c) rule-based methods: CP.

One early paper  reporting a solution to the AWS problem is Geiger's et al.
\cite{article8}.  They used the meta-heuristic TS with two additional
heuristics to minimize the $C_{max}$: (a) Profile Fitting (PF) heuristic and
(b) Nawaz-Enscore-Ham (NEH) heuristic. That paper reported  better results
when it is used NEH-TS. On the other hand, Bhushan and Karimi
\cite{article5} proposed a solution based on TS and SA, combined with new
specific algorithms for the same problem.
        
By contrast, there are only a handful of researchers reporting optimized
solutions to the AWS scheduling problem via a combinatorial approach. One of
them is due to Zeballos et al. \cite{article17} who reported a solution
using CP combined with different search strategies. Similarly, Novas and
Henning \cite{article11}reported another solution to the  problem using CP,
generalizing the problem by an innovative rolling horizon methodology and
taking into account the robot's empty movements.
        
ASP is a language for knowledge representation, based on non-monotonic logic
and designed to solve combinatorial problems \cite{article9}.  ASP has been
used to solve combinatorial knowledge-based problems in many different
areas. Among the most outstanding applications is the decision-making system
of a spacecraft \cite{article10}, planning for the generation of teams in a
seaport \cite{article12}, and autonomous vehicles in car assembly
\cite{article7}.
        
Despite the amazing characteristics of ASP, its grounding step represents an
important bottleneck in the solution process and we had to find a way to
reduce the number of combinations. To this end, we use the approach  of
exploiting information relative to the problem structure. In particular, we
propose a novel method based on ranges to reduce the size of the search
space. With this the performance and the size of the solved problems were
notoriously improved.

\section{Reducing the Search Space by a Method Based in Ranges}

The solving process in ASP consists of two stages: grounding and solving. It
has been recognized that  the grounding stage can represent a bottleneck
because of the number of combinations frequently builds huge domains. For
example, both in the planning and scheduling  it is necessary to include the
time. In this case, the resulting number of combinations becomes
prohibitively large, especially when  jobs require many processing units.
Therefore, for planning and scheduling problems, the grounding stage is a
highly demanding computational task and in the worst case, the task turns
intractable.

A common problem solving methodology in ASP is to architect the program in
three parts: GENERATE, DEFINE and TEST. The first part defines the
collection of answer sets that represent potential solutions. Auxiliary
concepts are defined in the DEFINE section. In our study case, this section
is used to define the problem constraints. The TEST section consists of
rules without head or constraints, used to eliminate answer sets that are
not solutions.

After numerous intents to solve the AWS scheduling problem with the classic
methodology used in ASP, it was evident that the problem could not be
solved, being the largest possible one of 4 baths and 4 jobs. Looking for a
more precise identification of the grounding trouble, the number of
combinations experimentally obtained is $N$, and depends on the number $N_b$
of baths, the number $N_j$ of jobs, and the deadline $D_d$, according to the
following  formula.

\begin{equation}
   N = Dd^{Nb*Nj}
\end{equation}

For example, for a problem with 8 baths and 8 jobs and deadline equal to
120, the number of combinations is 1.1684e+133.

As is evident, the number of combinations is huge and we have to approach
the problem in a different way. It is essential to find a way to cut down
the grounding process. So, we decided to reduce the size of the set of
possible solutions before implementing the generation rule. The application
of this method requires to be extremely careful on the heuristics' selection
that will be applied in order to reduce the number of combinations, because
it is possible that sets of answers containing the optimal solution could be
inadvertently deleted.

The selection of the heuristics is based on the knowledge of the problem and
its solutions. In particular, the heuristic  implemented is based on
generating combinations of possible solutions through ranges. As the problem
under study is about a flow workshop,  all jobs must be  processed
sequentially and all jobs must start on the first resource. Therefore, it is
useless to generate sequences with jobs starting its processing on the first
bath later than half the time available to process the complete job because
there would be insufficient time to finish the processing sequence. Since it
is unknown in advance which  jobs would be executed first, it was  decided
to do a formal modeling of the process. An advantage of this method is that
the solution space is greatly reduced. The model consists on calculating
ranges of sequences in which the solution is guaranteed to be found.

To do this, it is first necessary to calculate the sum of all processing
times for each bath. This sum must be done for each job, according to the
following equation:

\begin{equation}
   \lambda = \sum_{k=1}^{nb} t, \forall j
\end{equation}

where $n_b$ is the number of baths, $t$ is the processing time for each bath
and $j$ represents each job.
Later, $\lambda$ is used to find the maximum delay allowed for each job
according to the next formula

\begin{equation}
   \chi = \sum_{j=1}^{nj} deadline-\lambda_{j}
\end{equation}

where $n_j$ is the number of jobs and $\lambda_{j}$ is the total  processing
time for  job $j$.
Now, to calculate the range in which the combinations of each bath for a
given job must be generated, it is necessary to know the total processing
time from the first to the $i$-th bath, which is the previous bath whose
range is being calculated according to

\begin{equation}
   \rho = \sum_{k=1}^{th} t, \forall j
\end{equation}

where $th$ represents the $i$-th bath, and $t$ is the processing time in
each bath. Lastly, the superior and inferior limits of the range are
calculated using the values previously obtained, as shown next

\begin{equation}
   \Omega_{upper} = \chi_{j}+\rho_{j}, \forall j \forall k
\end{equation}

\begin{equation}
   \Omega_{lower} = \Omega_{{upper}_{j,k}} - \chi_{j} + 1 , \forall j
\forall k
\end{equation}

\section{Answer Set Programming Encoding}

Equations 2 to 6 are encoded in the following rules:

{\small\begin{lstlisting}
dBsumOneJ(J, L) :- job(W), J =1..W, L=#sum{D,B:  bath(M),B=1..M,
            job(W), J =1..W, duration(J,B,D)}.
maxDelayAnyJ(J,B,MaxR):- J=1..W, job(W), bath(M),B=1..M, dBsumOneJ(J,L),
            MaxR=deadline-L.
dSumJuntilB(J,Buntil,L) :- job(J), bath(Buntil),
            L=#sum{D,B:  bath(Buntil), B=1..Buntil, duration(J,B,D)}.
range(J,B,Ru,Rl) :- bath(B), job(J), duration(J,B,D),
            maxDelayAnyJ(J,B,MaxR), dSumJuntilB(J,B,L), Ru=MaxR+L,
            Rl=Ru-MaxR+1.
\end{lstlisting}}

The following is the generation rule, which calls  the previous rules that
calculates the ranges. In this way, a reduced version of the complete domain
is obtained.

{\small\begin{lstlisting}
{do(J,B,T,D) :- times(L), T=1..L, T+D<=deadline, range(J,B,Ru,Rl),
            T >= Rl, T <= Ru }=1:- duration(J,B,D), job(W),
            J=1..W, bath(M), B=1..M.
\end{lstlisting}}

Experimentally was found that applying the range method, the domain size is
reduced to 98.5\%. This percentage was calculated by first solving a $\pi_1$
program including only the generate rule. When $\pi_1$ program is solved,
$X$ number of models are found. Another $\pi_2$ program including  rules
related to equations 2 to 6 plus the generate rule was created. After the
$\pi_2$ program is executed by the solver, $Y$ number of models are found.
Dividing $Y$ by $X$ and multiplying the result by 100 the referred percent
is obtained. Given the huge size of  combinations, the experiment was only
tested on small problems.

The rules related to the resource uses were included In the \textbf{DEFINE}
section. The rules named \textbf{assigned} are used to decide if it is
possible to assign a job to some bath.  A job can be assigned to  bath $B_2$
if it was previously processed on $B_1$, and $B_2$ comes immediately after
$B_1$, and processing between $B_1$ and $B_2$ is not overlaped or ending
time is beyond of \textbf{deadline}. Rule \textbf{finished} is used to
determine if a job has finished its processing  in a certain bath.

{\small\begin{lstlisting}
assigned(J,B1,T1,D1):-  do(J,B1,T1,D1), do(J,B2,T2,D2), job(J), T1<T2,
            D=T1+D1, B2==B1+1, T2>=D, D <=deadline.
assigned(J,B2,T2,D2):- finished(J,B1,T1), do(J,B2,T2,D2), B2==B1+1,
                T2 >= T1, F=T2+D2, F<=deadline.
assigned(J2,B,T2,D2):- finished(J,B,T1), do(J2,B,T2,D2),
            not finished(J2,B,T3),J !=J2, times(T3),
                    T2 >= T1, F=T2+D2, T3 <=T2, F<=deadline.
 finished(J1,B1,T2):- assigned(J1,B1,T1,D1), T2=T1+D1, T
                    T1 < T2,  T2 <= deadline.

\end{lstlisting}}

We included a fourth section named \textbf{CONSTRAINTS} to the program. This
section includes constraints   such as: a) One of the scheduled jobs must start 
at bath one, rule \textbf{someJobStartAtOne}; b) baths must be processed sequentially, 
rule \textbf{anyJobStartSame}; c)  policies ZW/NIS for chemical baths and LS for water 
baths, rules \textbf{delayedBw}, \textbf{neg_delayedBw},  \textbf{busyBath} and 
\textbf{badSincronyBch}; d) rules to detect overlapping in resources use also were included 
in this section, such as \textbf{overlapBathJobs}, \textbf{overlapJobBaths} and 
\textbf{badStarting}; e) heuristic to avoid delay between a water bath and a chemical 
bath exceeding certain amount of time, rule \textbf{tooMuchDelay}, and; f) any job must 
finish after the deadline, rule \textbf{exceedsDeadline}. The rules in question were defined as: 

{\small\begin{lstlisting}
someJobStartAtOne(Mn):- Mn=#min{T1: times(T1), times(T2),
            do(J1,B,T1,D2), do(J2,B,T2,D1), B==1,T2 >T1}, Mn != 1.
anyJobStartSameT(T2):- do(J1,B,T1,D2), do(J2,B,T2,D1), J1
                    J1 != J2, T2 == T1.
delayedBw(J2,B2,T2,D2):-  finished(J2,B1,T1),
            assigned(J2,B2,T2,D2), B2==B1+1,  B1\2 == 0, T2>T1,
            job(JT), times(E), not neg_delayedBw(J2,B2,T2,D2).
delayedBw(F):-  finished(J1,B1,T1),
                    F=#min{T2 : finished(J1,B1,T1),
                    assigned(J2,B2,T2,D2), B2==B1+1, job(J2),
                    times(T2), bath(B2) }, F !=#sup, F > T1.
neg_delayedBw(J2,B2,T2,D2):-  finished(J2,B1,T1),
                assigned(J2,B2,T2,D2), B2==B1+1,  B1\2 == 0,
                    T2>T1, job(JT), times(E), not delayedBw(J2,B2,T2,D2).
neg_delayedBw(J2,B2,T2,D2):- busyBath(J2,B2,T2,D2).
busyBath(J2,B2,T2,D2):-         finished(J2,B1,T1),
                    assigned(J2,B2,T2,D2), job(JT), JT != J2,
                    assigned(JT,B2,T3,D3), T2>T1,  B1\2 == 0,
                    B2==B1+1, F=#max{E: finished(J2,B1,T1),
                    assigned(J2,B2,T2,D2), job(JT),
                    JT != J2, assigned(JT,B2,T3,D3),
                    E=T3+D3, finished(JT,B2,E)},
                    T3 <=T1, F !=#inf, F==T2,F>T1.

badSincronyBch(T2):- do(J1,B1,T1,D1), do(J1,B2,T2,D2),
                    B1\2 > 0, B2 ==B1+1, T2 > T1+D1.

overlapBathJobs(T2):- do(J1,B,T1,D1), do(J2,B,T2,D2),
                    bath(B), duration(J1,B,D1), J1 < J2, T1 <= T2,
                    T1 + D1 > T2.
overlapJobBaths(T2):- do(J1,B1,T1,D1), do(J1,B2,T2,D2), B2
                    B2 == B1+1, T1<=T2, T1+D1 > T2.
badStarting(T2):- do(J,B2,T2,D2), do(J,B1,T1,D1),
                    B2 > B1, T2 < T1+D1.
tooMuchDelay(J,B1,B2,F1,T2):- do(J,B1,T1,D1),
                do(J,B2,T2,D2), B1\2==0, B2==B1+1, F1=T1+D1,
            T2>F1, maxDur(M), L=T2-F1, L>2
exceedsDeadline(Tf):- do(J1,B1,T1,D1), do(J2,B2,T2,D2),
                    T2 >=T1, Tf=T2+D2, Tf > deadline.
\end{lstlisting}}

\section{Experimental Results}
The data used in this research were artificially created and published by
Bhushan and Karimi \cite{article4}. This data has also been used by other
researchers, such as \cite{article17} \cite{article2}. However, it was not
possible use this data directly  in ASP, because it contains real values
and ASP only works with integers. Thus, the original data was multiplied by
ten. After the problem was solved by ASP, the  solution values were divided
again by ten. The results shown in the next table were obtained through the
experiment with and without the method based in ranges. In this experiment
we used a  small computer with an Intel processor Core i5.3337 CPU@1.80 Ghz
X4, 6 Gb of memory, and 64 bits Ubuntu. The ASP solver used was clingo 4.4.

\begin{table}
\caption{Results obtained with a typical rule and the range method proposed
\newline time is in seconds.  NS: No solution}
\label{tab:SolnsMaq1}
\begin{center}
\begin{tabular}{| c | c | c | c |}
\hline
P:[BxJ] & makespan & CPU Time & Approach \\ \hline
[2x2]  & 18 & 1.820 & No ranges \\ \hline
[2x2]  & 18 & 0.030 & with ranges \\ \hline
[4x4]  & 50 & 96.920 & No ranges \\ \hline
[4x4] & 50 & 11.250 & with ranges \\ \hline
[5x5] & -- & NS & No ranges \\ \hline
[5x5] & 65 & 165.890 & with ranges \\ \hline
\end{tabular}
\end{center}
\end{table}

To test the hardware constraints on the solving process performance, the
model based in ranges for problems of  6 baths with 6 jobs and 7 baths with
7 jobs was tested on a  computer with CPU @2.88 Ghz, 8 cores, and memory of
16 Gigabytes. Only an optimal solution was found for the 6 baths - 6 jobs
problem in a CPU time of 240.280 seconds and a makespan of 83. Neither the
optimal solution nor the first solution was found.

\section{Conclusions}
ASP solver’s limitation to solve big scheduling and planning problems comes
from the approach used in the grounding step because a huge search space
needs to be previously created for the reasoner to start looking for the
solution. Although this problem is mitigated in some way by different
strategies applied both in the grounding and solving stages, they are not
effective enough in problems whose characteristics include  variables like
time, which turn the problem into one whose input grows exponentially.
In our study case the product is processed sequentially making it convenient
to exploit its structure to avoid generate useless combinations, like those
starting at a certain time that  would turn impossible to complete the
processing. The proposed method only creates combinations with possibilities
to finish on time and allowed to reduce the search space up to 98.5\%.
Although this reduction is important, it only allowed to increase the
problem size from 4 bathrooms and 4 jobs to 6 bathrooms and 6 jobs. This
results are not so surprising if we consider the example  for a deadline of
83, with 6 jobs and 6 bathrooms, where the number of combinations was
8.0995e + 96. Using the range method the problem’s size will still have
1.2149e + 95 combinations. 

Obviously, these kinds of problems need to be approached through a powerful
methodology using a different strategy in the grounding step. It is foreseen
that better results can be obtained approaching the problem with Constraint
Answer Set Programming (CASP), because this hybrid methodology uses a
completely different strategy in the grounding step. However, this approach
is motive of anoher research what would be published in the future.
Nonetheless, we consider that the proposed method is applicable to small
flowshop scheduling problems as well as to establish a way to exploit the
problem structure in order to reduce the number of possible solutions and
improve the solver performance.
\bibliographystyle{eptcs}
\bibliography{BibRanges}
\end{document}